\documentstyle [12pt] {article}

\begin{document}



\newcount\itemnum  
\def\newitem{\itemnum=0}
\def\nextitemsup{\advance\itemnum by1 \item{$^{\the\itemnum}$}}
\def\nextitemp{\advance\itemnum by1 \item{{\the\itemnum}.}}

\newcount\refnum 
\def\newref{\refnum=0}
\newref

\newcount\eqnum 
\def\neweq{\eqnum=0}
\neweq

\newcount\istyle 
\def\stylepr{\istyle=1} 
\def\stylenp{\istyle=2} 
\def\styleap{\istyle=3} 

\newcount\localref 
\def\newloc{\localref=0}
\newloc

\def\nextref#1{\advance\refnum by1
    \header\outref{\the\refnum}\trailer
    \edef#1{\the\refnum}}
\def\reff#1{\header\outref{#1}\trailer} 
\def\nextrefs#1{\advance\refnum by1 \header
    \outref{\the\refnum,\hskip -.41em plus .08em}\edef#1{\the\refnum}
    \advance\localref by1}
\def\refs#1{\header \outref{#1,\hskip -.41em plus .08em}
    \advance\localref by1}
\def\outref#1{\ifnum\istyle=3{#1}\else{$^{#1}$}\fi}
\def\header{\ifnum\localref=0{\ifnum\istyle=3{[}\fi}\fi}
\def\trailer{\newloc \ifnum\istyle=2{$^)$}\fi \ifnum\istyle=3{]}\fi}

\newcount\secnum 
\def\newsec{\secnum=0}
\newsec

\newcount\subsecnum 
\def\newsubsec{\subsecnum=0}
\newsubsec

\def\section#1{\advance\secnum by1 \outsec{\the\secnum}{#1}
\newsubsec
     \ifnum\istyle=2\neweq\fi}
\def\outsec#1#2{\ifnum\istyle=2 \centerline{#1. #2}
     \else \centerline{\uppercase\expandafter{\romannumeral #1}.
     \uppercase{#2}} \fi}

\newcount\locoffset 

\def\numlet#1{\locoffset=96 \advance\locoffset by #1
     \char\locoffset}
\def\Numlet#1{\locoffset=64 \advance\locoffset by #1
     \char\locoffset}

\def\subsection#1{\advance\subsecnum by1 \vskip 0.6truein
\outsubsec{#1}}
\def\outsubsec#1{\ifnum\istyle=2
\leftline{\the\secnum.\the\subsecnum\ \  #1}
     \else\centerline{\Numlet{\subsecnum}.\  #1}\fi}

\def\pagenumbers{\footline={\hss\tenrm\folio\hss}}

\def\nexteq{\global\advance\eqnum by1
     \ifnum\istyle=
2{(\the\secnum}.{\the\eqnum})\else{(\the\eqnum})\fi}

\newcount\subeqlet 
\def\nexteqp{\global\advance\eqnum by1 \global\subeqlet=1 \outeqp}
\def\sameeq{\global\advance\subeqlet by1 \outeqp}
\def\outeqp{\ifnum
     \istyle=2{(\the\secnum}.{\the\eqnum}\numlet{\subeqlet})%
     \else{(\the\eqnum}\numlet{\subeqlet})\fi}

\stylenp        


\newcommand {\ket} [1] { \left| \, #1  \right> }

\newcommand {\bra} [1] { \left<  #1 \, \right| }

\newcommand {\ph} [1] { (-1)^{#1} }

\newcommand {\cgcsimp} [2] { \cgc{p_{#1},q_{#1};k_{#1},l_{#1},m_{#1}}
{p_{#2},q_{#2};k_{#2},l_{#2},m_{#2}} {p,q;k,l,m} }

\newcommand {\wcg} [7]
{   C   \begin{array}{ccc}{\scriptstyle [#1]} & {\scriptstyle
 \stackrel{#4}{[#2]} } & {\scriptstyle [#3]} \\
       {\scriptstyle [#5]}  & {\scriptstyle [#6] } & {\scriptstyle
 [#7] } \end{array}   }

\newcommand {\cgc} [6]
{   C   \begin{array}{ccc}{\scriptstyle #1} & {\scriptstyle #2} &
 {\scriptstyle #3} \\
       {\scriptstyle #4}  & {\scriptstyle #5 } & {\scriptstyle #6 }
 \end{array}  }


\begin{center}
{\large\bf  SU3 isoscalar factors}

\bigskip
{\it H. T. Williams} \\
\smallskip
Department of Physics, Washington and Lee University, \\
Lexington, VA 24450, USA \\
September, 1995 \\
\bigskip

\medskip\medskip
{\bf Abstract}
\medskip
\end{center}

\noindent A summary of the properties of the Wigner
Clebsch-Gordan coefficients and isoscalar factors for
the group SU3 in the SU2$\otimes$U1 decomposition is presented.
The outer degeneracy problem is discussed in detail with a proof
of a conjecture (Braunschweig's) which has been the basis of previous
work on the SU3
coupling coefficients.  Recursion relations obeyed by the SU3 isoscalar
factors are produced, along with an algorithm which allows numerical
determination of the factors from the recursion
 relations.  The algorithm produces isoscalar factors which share all
the symmetry properties under permutation of states and conjugation
which are familiar from the SU2 case.  The full set of symmetry
properties for the SU3 Wigner-Clebsch-Gordan coefficients and
isoscalar factors are displayed.

\clearpage

{\flushleft {\bf I. Introduction}}
\medskip

The group SU3 continues to be useful in modeling symmetries observed in
particle and nuclear physics.  In the late 1950's it found application
in classification of ``elementary'' hadrons, and in the description of
rotational states of non-spherical nuclei.
 Its utility persists as the color symmetry of quantum chromodynamics,
various models for collective nuclear motion, and elsewhere.

The Wigner-Clebsch-Gordan coefficients (WCG) are of particular interest.
These can be defined as the expansion coefficients of a composite state
of good SU3 quantum numbers in terms of direct products of two individual
SU3 classified states, paralleling
 Wigner's original use of SU2 in the treatment of quantum angular momentum.
The WCG can also be developed as the matrix elements of a set of tensor
operators which have distinctive properties under the transformations of
SU3.  These two viewpoints on the
 WCG are formally identical, and their algebraic connection is expressed
by the Wigner-Eckart Theorem.  The SU3 case presents a complication that
is absent in the SU2 recoupling problem -- that of the {\it outer
degeneracy.}  The complete determination of
 WCG's in SU3 requires a criterion outside the SU3 group to completely
classify composite states, and thus to fully define numerical values
for the WCG.  In {\it this} process, the two perspectives on the WCG
mentioned above suggest quite different mechanisms.

Biedenharn and coworkers \cite{bied1,bied2,bied3}, adopting the operator
point of view, have developed a set of canonical SU3-labeled unit tensor
operators, whose matrix elements become the ``canonical WCG's.''  The
canonical operators acquire SU3 labels
by virtue of their behavior under the transformations of the group.  As
well, each produces a unique set of shifts -- i.e, its action when
operating on  a state from a particular irreducible representations
(irrep) produces states from a unique second irrep;
and in cases of non-trivial outer degeneracy there is a distinct
operator for each degeneracy index.  The uniqueness of the operators,
and thus their designation as canonical, comes from their null space
properties.  The {\it characteristic null space}
of an operator is the union of all irreps which identically yield
zero under the action of the operator.  In the case of a tensor operator
of degeneracy one, the null space is uniquely determined by the group
properties of the operator and the state operated
upon:  for higher degeneracy the operators for distinct degeneracy
labels are chosen to have a null space each larger than the previous and
completely containing it.

   Adopting the alternative viewpoint -- the WCG's as coupling coefficients
which give the amplitude for the joining of two SU3 states to a composite
state of good SU3 quantum numbers -- coefficients which exhibit symmetry
under interchange of the two states
being coupled are suggested (a symmetry missing from the canonical
coefficients \cite{bied4}.)  WCG's for those couplings which have
degeneracy 1 possess this symmetry, and indeed all the symmetries under
permutation of irreps of the familiar SU2 Clebsch-Gordan
coefficients.   It has been proven that such permutation
symmetric WCG's for the SU3 case with degeneracy $> 1$ exist
\cite{sym1,sym2,sym3} and examples of such SU3 WCG's have been
developed \cite{sym4,sym5}.

In the sections which follow, new results pertaining to the SU3 WCG's
are presented which simplify evaluation of the WCG's, and which are
independent of the particular scheme adopted for outer degeneracy
resolution.  In particular, a collection of recursion
relations are defined for the isoscalar factors.  An algorithm
is presented which utilizes these recursion relations to generate
a set of WCG's demonstrating all the Racah symmetries familiar from
SU2.  This algorithm has been used in a successful C language implementation.

\medskip\medskip
{\flushleft {\bf II. Definitions and Notation}}
\medskip

A linear vector space which carries an irrep of SU3 is fully specified by two
integers $(p,q)$ henceforth referred to as the {\it irrep labels.}  The
dimension of the space is
\begin{equation}
 d = (p+1)(q+1)(p+q+2)/2. \label{dimens}
\end{equation}
A complete set of $d$ orthogonal vectors within the irrep can be labeled
by three further integers $(k,l,m)$, the {\it subspace labels,} which
satisfy the {\it betweenness conditions}
\begin{equation}
   p+q \geq k \geq q \geq l \geq 0 \mbox{ ; } k \geq m \geq l.
    \label{betwee}
\end{equation}
A fully specified member of the orthonormal spanning set for the
irrep is denoted by the ket
\[ \ket{p,q;k,l,m}. \]

The WCG are the coefficients ($C$) of the expansion of a composite
SU3 state ket in terms of products of SU3 kets
 \begin{equation} \ket{ {\cal P}; \kappa } = \sum
  \wcg{{\cal P}_1}{{\cal P}_2}{{\cal P}}{n}{\kappa_1}{\kappa_2}{\kappa} \,
  \ket{{\cal P}_1;  \kappa_1} \, \ket{{\cal P}_2;  \kappa_2} . \label{wcgdef}
  \end{equation}
where ${\cal P}$ is shorthand for the pair $(p,q)$ and $\kappa$ for the set
$(k,l,m)$; the sum extends over subspace labels of $\kappa_1$ and $\kappa_2$;
and $n = 0,1, \ldots$ labels the outer degeneracy.  The Wigner-Eckart Theorem
relates the WCG's to matrix
elements of operators $T^n_{p,q;k,l,m}$ which transform like tensors
under the operations of SU3
\begin{eqnarray}
 \bra{p,q;k,l,m} T^n_{p_1,q_1;k_1,l_1,m_1} \ket{p_2,q_2;k_2,l_2,m_2} =
     \nonumber \\
\wcg{{\cal P}_1}{{\cal P}_2}{{\cal P}}{n}{\kappa_1}{\kappa_2}{\kappa} \; <
  p,q \, \| T^n_{p_1,q_1} \| \, p_2,q_2 >,
\end{eqnarray}
where
\[ < \, p,q \, \| T_{p_1,q_1} \| \, p_2,q_2 \, >,  \]
the {\it reduced matrix element,} is a complex number which depends only
upon the three sets of irrep labels. (The unit tensor operators of the
Biedenharn scheme are so named since each has a reduced matrix element
of one.)

The ket labeling scheme described above represents the decomposition
$SU3 \supset SU2 \otimes U1.$  The labels $(k,l,m)$ are the quantum
numbers of the $SU2$ subgroup, and are related to the isospin ($I$)
and its $z$ component ($I_z$) by
\begin{eqnarray}
   I & = & \frac{k-l}{2} \\
   I_z & = & m - \frac{k+l}{2}; \label{subspI}
\end{eqnarray}
and the $U1$ subgroup with the hypercharge ($Y$) given by
\begin{equation}
   Y = k+l-\frac23 (p+2q).   \label{subspY}
\end{equation}
The WCG of equation~(\ref{wcgdef}) will vanish unless the subspace
labels obey the relations
\begin{eqnarray}
   I_1 + I_2 & \geq & I,   \nonumber \\
   | I_1 - I_2 | & \leq & I, \label{triang} \\
   I_{1_z} + I_{2_z} & = & I_z,  \label{zcompI} \\
   Y_1 + Y_2 & = & Y.  \label{hyperc}
\end{eqnarray}
This decomposition allows factoring of an SU2 Clebsch-Gordan coefficient
from the SU3 WCG as follows:
\begin{equation}
    \wcg{{\cal P}_1}{{\cal P}_2}{{\cal P}}{n} {\kappa_1}{\kappa_2}{\kappa}
 \, =   \cgc{I_1}{I_2}{I}{I_{1_z}}{I_{2_z}}{I_z} \,
 F^n(p,q,k,l:p_1,q_1,k_1,l_1;p_2,q_2,k_2,l_2) \label{ISFdef}
\end{equation}
where the factor $F$, which is independent of the $m$ subspace labels, is
called the {\it isoscalar factor} (ISF). In subsequent usage, when their
values are obvious from the context, the $p$ and $q$ values will be
suppressed in the notation for the ISF.

A particular set of subspace labels, $k=m=p+q$, $l=0$, will play an
important role in the present consideration.  This set, referred to as
the {\it state of highest weight} for a particular irrep, will be
referred to by the replacement $(k=p+q,l=0,m=p+q)
\rightarrow SHW$ and likewise in the isoscalar factor by $(p,q,k=p+q,l=0)
 \rightarrow (p,q,SHW).$

\medskip\medskip
{\flushleft {\bf III. Outer degeneracy}}
\medskip

The Clebsch-Gordan series for SU3
\begin{equation}
    (p_1,q_1) \otimes (p_2,q_2) = \sum_i {\eta}_i \,(p_{i}^{'},q_{i}^{'})
\end{equation}
indicates the number of distinct times (${\eta}_i$) the irrep
$(p_{i}^{'},q_{i}^{'})$ appears in the outer product of irreps
$(p_1,q_1)$ and $(p_2,q_2)$.  The circumstance of ${\eta}_i > 1$
is a feature of $SU3$ referred to as {\it outer degeneracy}, and
the coefficients ${\eta}_i$ will be referred to herein as the
{\it degeneracy} of the coupling $(p_1,q_1) \otimes (p_2,q_2)
 \rightarrow  (p_{i}^{'},q_{i}^{'}).$

The value of the degeneracy is a function of the six
 irrep labels and can be deduced from the betweenness
 conditions of equation~(\ref{betwee}) for each irrep,
 and the requirements of the $SU2$ and $U1$ subgroups,
 given in equations~(\ref{triang}), (\ref{zcompI}) and
 (\ref{hyperc}).  These latter requirements
 follow from the corresponding Clebsch-Gordan series for $SU2$
(triangularity of three Euclidian vectors in two dimensions) and
for $U1$ (scalar addition.)  Various ways of evaluating the degeneracy appear
in the literature \cite{oreil,bied5}.  An equivalent expression for
the degeneracy consistent with present notation is
\begin{eqnarray}
   {\eta} & =  max( & \! \! {\eta}^{\prime} +1-max(\gamma,\sigma),0)
\nonumber \\
   {\eta}^{\prime}  & =   min( & \! \! p_1+\sigma,p_2+\sigma,q+\sigma,
q_1+\gamma,q_2+\gamma,p+\gamma,2(\sigma+\gamma), \nonumber \\ & &
 p_1+q_1-\gamma-\sigma,p_2+q_2-\gamma-\sigma) \label{degene}
\end{eqnarray}
where $\gamma \equiv (p_1+p_2-p)/3,$ and $\sigma \equiv (q_1+q_2-q)/3.$

This expression can be used to prove Braunschweig's conjecture which has
been used by several authors \cite{braun,leblan,bied6} in work related to
 determination of WCG's for SU3.  The conjecture suggests that the number
 of non-vanishing values of the WCG

\begin{equation}
   \wcg{{\cal P}_1 }{{\cal P}_2}{{\cal P }}{n} {SHW}{\kappa_2}{SHW}
  \label{shwwcg}
\end{equation}
is no less than the degeneracy of the irrep coupling.
The value for the subspace label $m_2$ is fixed by
equation~(\ref{zcompI}) and $l_2$ is dependent upon
$k_2$ through equation~(\ref{hyperc})
\begin{equation}
   k_2 + l_2 = p_2 + q_2 - \gamma + \sigma , \label{proof1}
\end{equation}
so counting the number of non-vanishing WCG's of this
type can be accomplished by determining the range of $k_2$ values.
Upper and lower limits on $k_2$ come from the triangularity expressions
of equation~(\ref{triang}) combined with equation~(\ref{proof1}), producing
\begin{eqnarray}
   k2 & \geq & p_2+q_2 - 2 \gamma - \sigma \nonumber \\
   k_2 & \geq & \gamma + 2 \sigma \nonumber \\
   k_2 & \leq & p + q + \gamma + 2 \sigma .
\end{eqnarray}
The betweenness relations for state 2 give further limits on $k_2$
($p_2+q_2 \geq k_2 \geq q_2$) and on $l_2$ ($q_2 \geq l_2 \geq 0$),
which when combined with equation (\ref{proof1}) produce
\begin{eqnarray}
   k_2 & \geq & \sigma - \gamma +p_2, \nonumber \\
   k_2 & \leq & \sigma - \gamma + p_2 + q_2.
\end{eqnarray}
The limits on $k_2$ are thus
\begin{equation}
   max(p_2+q_2-2 \gamma - \sigma, \gamma + 2 \sigma, q_2, p_2 +
 \sigma - \gamma) \leq k_2 \leq min(p_2+q_2+ \sigma - \gamma,
 p_2+q_2, p + q + \gamma + 2 \sigma)
\end{equation}
which implies that the total number of $k_2$ values producing
non-vanishing WCG's in this case is given by
\begin{eqnarray}
   1 + min& \!\! (& \!\! \gamma + 2 \sigma, \sigma + 2 \gamma,
 p_2+q_2 - \sigma - 2 \gamma, p_2+q_2 - \gamma - 2 \sigma, q_2, \nonumber \\
   & & q_2+\gamma - \sigma, p_1+q_1, p+q, p+q-q_2+\gamma + 2 \sigma,
 p+q-p_2+2 \gamma + \sigma).
\end{eqnarray}
A term-by-term comparison of this expression with that for the
degeneracy (equation~(\ref{degene})) reveals that the number of
$k_2$ values producing non-vanishing WCG's of the form of
equation~(\ref{shwwcg}) is greater than or equal to the
degeneracy.
This inequality is sufficient for the uses of the previously
cited references, and has been important in the development
of the algorithm presented later in this paper.

\medskip\medskip
{\flushleft {\bf III. Recursion relations}}
\medskip

An efficient scheme for evaluation of the SU3 ISF {\it independent of the
method chosen for resolution of outer degeneracy} involves use of recursion
relations for these quantities, which can be derived using the group
generators \cite{GlasGas}.  Explicit
 expressions for the generators depend upon a choice of signs of the
matrix elements of the generators between elements of the fundamental three-
dimensional representation:  those given below follow the phase convention
of de Swart \cite{deS}. The actions
of these generators on the orthonormal kets previously defined are
\begin{eqnarray}
	{\hat T}_+ \, | \, p,q; \, k,l,m \, > & = & \sqrt{(k-m)(m-l+1)} \,
	       | \, p,q; \, k,l,m+1 \, > \label{tplus} \\
	{\hat T}_- \, | \, p,q; \, k,l,m \, > & = & \sqrt{(k-m+1)(m-l)} \,
	       | \, p,q; \, k,l,m-1 \, > \label{tminus} \\
	{\hat V}_+ \, | \, p,q; \, k,l,m \, > & = &
	     \sqrt{\frac{(k+2)(m-l+1)(k-q+1)(p+q-k)}{(k-l+1)(k-l+2)}} \,
	       | \, p,q; \, k+1,l,m+1 \, > \nonumber \\
	 & & + \sqrt{\frac{(l+1)(k-m)(q-l)(p+q-l+1)}{(k-l)(k-l+1)}} \,
	       | \, p,q; \, k,l+1,m+1 \, > \label{vplus} \\
	{\hat V}_- \, | \, p,q; \, k,l,m \, > & = &
	     \sqrt{\frac{(k+1)(m-l)(k-q)(p+q-k+1)}{(k-l)(k-l+1)}} \,
	       | \, p,q; \, k-1,l,m-1 \, > \nonumber \\
	 & & + \sqrt{\frac{l(k-m+1)(q-l+1)(p+q-l+2)}{(k-l+1)(k-l+2)}} \,
	       | \, p,q; \, k,l-1,m-1 \, > \label{vminus} \\
	{\hat U}_+ \, | \, p,q; \, k,l,m \, > & = &
	     \sqrt{\frac{(k+2)(k-m+1)(k-q+1)(p+q-k)}{(k-l+1)(k-l+2)}} \,
	       | \, p,q; \, k+1,l,m \, > \nonumber \\
	 & & - \sqrt{\frac{(m-l)(l+1)(q-l)(p+q-l+1)}{(k-l)(k-l+1)}} \,
	       | \, p,q; \, k,l+1,m \, > \label{uplus} \\
	{\hat U}_- \, | \, p,q; \, k,l,m \, > & = &
	     \sqrt{\frac{(k+1)(k-m)(k-q)(p+q-k+1)}{(k-l)(k-l+1)}} \,
	       | \, p,q; \, k-1,l,m \, > \nonumber \\
	 & & - \sqrt{\frac{l(m-l+1)(q-l+1)(p+q-l+2)}{(k-l+1)(k-l+2)}} \,
	       | \, p,q; \, k,l-1,m \, > \label{uminus}.
\end{eqnarray}
Three diagonal operators indicate the values of the subspace labels for a ket:
\begin{eqnarray}
	{\hat T}_3 \, | \, p,q; \, k,l,m \, > & = & I_z \,
	       | \, p,q; \, k,l,m \, > \label{tthree} \\
	{\hat Y} \, | \, p,q; \, k,l,m \, > & = & Y \,
	       | \, p,q; \, k,l,m \, > \label{Yop} \\
	{\hat T}^2 \, | \, p,q; \, k,l,m \, > & = & \frac12 ( {\hat T}_+
	       {\hat T}_- + {\hat T}_- {\hat T}_+ )| \, p,q; \, k,l,m \, >
	       \nonumber \\
	 & = & I(I+1)  \,
	       | \, p,q; \, k,l,m \, > \label{Isqop}
\end{eqnarray}
making use of the definitions of equations~(\ref{subspI}) and
(\ref{subspY}).  The operators ${\hat T}_+$ and ${\hat T}_-$ move up and
down in the variable $I_z$, and thus have no effect on the ISF.
The remaining four nondiagonal (ladder) operators form
the basis of the derivation of the recursion relations.

Consider a composite state of highest weight
 \begin{equation} \ket{ {\cal P}; SHW } = \sum
  \wcg{{\cal P}_1}{{\cal P}_2\;\;}{{\cal P}\,\,\,\,\,\,}{n}{\kappa_1 \,}
		    {\kappa_2 \,}{SHW} \,
  \ket{{\cal P}_1;  \kappa_1} \, \ket{{\cal P}_2;  \kappa_2} .   \label{shwcgc}
\end{equation}
The action of ${\hat V}_+$ on this ket must vanish since each of the two
states it produces have $m = p+q+1$ which violates betweenness.  Linearity
of the generators (e.g. ${\hat V}_+ = {\hat V}_{1+} + {\hat V}_{2+}$) implies
from equation~(\ref{shwcgc}) that
\begin{eqnarray}
    {\hat V}_+ \ket{ {\cal P}; SHW } & = & 0 \nonumber \\
   & = & \sum \wcg{{\cal P}_1}{{\cal P}_2}{{\cal P}}{n}{\kappa_1 }{\kappa_2 }
	{SHW} \left( \ket{{\cal P}_2;  \kappa_2} {\hat V}_{1+}\ket{{\cal P}_1;
      \kappa_1} + \ket{{\cal P}_1;  \kappa_1} {\hat V}_{2+} \ket{{\cal P}_2;
      \kappa_2} \right).
\end{eqnarray}
Use of the defining equation for ${\hat V}_+$ (equation~(\ref{vplus})) changes
this expression into a summed four-term expression which must vanish.  The
orthogonality of any two SU3 kets with different subspace labels allows this
sum to be transformed into a four term recursion relation for the WCG:
\begin{eqnarray}
   0 & = & \sqrt{ \frac{(k_1+1)(m_1-l_1)(k_1-q_1)(p_1+q_1-k_1+1)}
				  {(k_1-l_1)(k_1-l_1+1)} } \;
    \wcg{{\cal P}_1 }{{\cal P}_2 }{{\cal P}}{n}{k_1-1,l_1,m_1-1}
	   {k_2,l_2,m_2}{SHW} \nonumber \\ & + &
  \sqrt{ \frac{(k_2+1)(k_2-q_2)(p_2+q_2-k_2+1)(m_2-l_2+1)}
	       {(k_2-l_2)(k_2-l_2+1)} } \;
    \wcg{{\cal P}_1 }{ {\cal P}_2 }{{\cal P}}{n}{k_1,l_1,m_1}
       {k_2-1,l_2,m_2-1}{SHW} \nonumber \\ & + &  \sqrt { \frac
	{l_1(q_1-l_1+1)(k_1-m_1+1)(p_1+q_1-l_1+2)}
	   {(k_1-l_1+1)(k_1-l_1+2)} } \;
    \wcg{{\cal P}_1}{{\cal P}_2 }{ {\cal P}}{n}{k_1,l_1-1,m_1-1}
		{k_2,l_2,m_2}{SHW}
\nonumber \\ & + &  \sqrt{  \frac{l_2(k_2-m_2+1)(q_2-l_2+1)(p_2+q_2-l_2+2)}
	      {(k_2-l_2+1)(k_2-l_2+2)}  } \;
    \wcg{{\cal P}_1, }{ {\cal P}_2 }{{\cal P}}{n}{k_1,l_1,m_1}{k_2,l_2-1,m_2-1}
	   {SHW}
\end{eqnarray}
This should be valid for any set of projection quantum numbers,
thus for $k_1=m_1$ which allows the replacement of the WCG in this
expression by products of ISF and simple SU2 Clebsch-Gordan coefficients
whose values can be expressed analytically \cite{varsh}.
The result of this replacement is a four term recursion relation
among the ISF's for coupling to a state of highest weight:
\begin{eqnarray*}
0 & = &  a_1 F^n(SHW:k_1-1,l_1;k_2,l_2) + a_2 F^n(SHW:k_1,l_1;k_2-1,l_2), \\
 & & - a_3 F^n(SHW:k_1,l_1-1;k_2,l_2) +a_4 F^n(SHW:k_1,l_1;k_2,l_2-1),
\end{eqnarray*}
where
\begin{eqnarray}
 a_1 & = & \sqrt{\frac{(k_1+1)(k_1-q_1)(p_1+q_1-k_1+1)(p+q+2I_1+2I_2+3)
   (p+q+2I_1-2I_2+1)}{I_1(2I_1+1)}},\nonumber  \\
 a_2 &=& \sqrt{\frac{(k_2+1)(k_2-q_2)(p_2+q_2-k_2+1)(p+q+2I_1+2I_2+3)
   (p+q-2I_1+2I_2+1)}{I_2(2I_2+1)}}, \nonumber \\
 a_3 &=& \sqrt{\frac{l_1(q_1-l_1+1)(p_1+q_1-l_1+2)(-p-q+2I_1+2I_2+1)
   (p+q-2I_1+2I_2+1)}{(2I_1+1)(I_1+1)}}, \nonumber \\
 a_4 &=&  \sqrt{\frac{l_2(q_2-l_2+1)(p_2+q_2-l_2+2)(-p-q+2I_1+2I_2+1)
   (p+q+2I_1-2I_2+1)}{(2I_2+1)(I_2+1)}}. \label{RR1}
\end{eqnarray}

Similarly,
\begin{equation}
   {\hat U}_- \, \ket{ {\cal P}; SHW } = 0
\end{equation}
since the two kets produced by this operation both violate betweenness.
By an analogous set of steps one derives a second, distinct recursion relation:
\begin{eqnarray*}
0 & = &  b_1 F(SHW:k_1+1,l_1;k_2,l_2) - b_2 F(SHW:k_1,l_1;k_2+1,l_2), \\
 & & + b_3 F(SHW:k_1,l_1+1;k_2,l_2) +b_4 F(SHW:k_1,l_1;k_2,l_2+1),
\end{eqnarray*}
where
\begin{eqnarray}
 b_1 & = & \sqrt{\frac{(k_1+2)(k_1-q_1+1)(p_1+q_1-k_1)(-p-q+2I_1+2I_2+1)
   (p+q-2I_1+2I_2+1)}{(2I_1+1)(I_1+1)}}, \nonumber \\
 b_2 & = & \sqrt{\frac{(k_2+2)(k_2-q_2+1)(p_2+q_2-k_2)(-p-q+2I_1+2I_2+1)
   (p+q+2I_1-2I_2+1)}{(2I_2+1)(I_2+1)}}, \nonumber \\
 b_3 & = & \sqrt{\frac{(l_1+1)(q_1-l_1)(p_1+q_1-l_1+1)(p+q+2I_1+2I_2+3)
   (p+q+2I_1-2I_2+1)}{I_1(2I_1+1)}}, \nonumber \\
 b_4 & = &  \sqrt{\frac{(l_2+1)(q_2-l_2)(p_2+q_2-l_2+1)(p+q+2I_1+2I_2+3)
   (p+q-2I_1+2I_2+1)}{I_2(2I_2+1)}}.  \label{RR2}
\end{eqnarray}

One can "step down" from the ISF's for the coupled state of highest weight
to ISF's for any other $k,l$ values by use of two relations derived in an
analogous fashion from the actions of the operators ${\hat V}_-$ and
${\hat U}_+$, respectively:
\begin{eqnarray}
F^n(k,l:k_1,l_1;k_2,l_2)  =  \alpha ( c_1 F^n(k+1,l-1:k_1,l_1;k_2,l_2)
 + c_2 F^n(k,l-1:k_1,l_1-1;k_2,l_2) \nonumber \\ - c_3
 F^n(k,l-1:k_1,l_1;k_2-1,l_2) +
 c_4 F^n(k,l-1:k_1,l_1;k_2,l_2-1) ) \label{RR3}
\end{eqnarray}

where
\begin{eqnarray*}
 \alpha & = & \frac{k-l+2}{\sqrt{2l(q-l+1)(p+q-l+2)}} \\
 c_1   & = & \sqrt{\frac{(k+2)(k-q+1)(p+q-k)(I_1+I_2-I)
(-I_1+I_2+I+1)}{(I+1)^2(I_1+I_2+I+2)(I_1-I_2+I+1)}} \\
c_2    & = & \sqrt{\frac{4 l_1 (q_1-l_1+1)(p_1+q_1-l_1+2)(2I_1+1)}{
(2I_1+2)(2I_1+2I_2+2I+4)(2I_1-2I_2+2I+2)}} \\
c_3    & = & \sqrt{\frac{(k_2+1)(k_2-q_2)(p_2+q_2-k_2+1)(I_1+I_2-I)}
{I_2(2I_2+1)(I_1-I_2+I+1)}} \\
c_4    & = & \sqrt{\frac{l_2(q_2-l_2+1)(p_2+q_2-l_2+2)(-I_1+I_2+I+1)}
{(2I_2+1)(I_2+1)(I_1+I_2+I+2)}};
\end{eqnarray*}
and
\begin{eqnarray} F^n(k,0:k_1,l_1;k_2,l_2) & = & \beta
 ( d_1 F^n(k+1,0:k_1+1,l_1;k_2,l_2) \nonumber \\
&& + d_2 F^n(k+1,0:k_1,l_1;k_2+1,l_2)+ d_3 F^n(k+1,0:k_1,l_1;k_2,l_2+1) )
 \label{RR4}
\end{eqnarray}
where
\begin{eqnarray*}
 \beta & = & \sqrt{\frac{(k+2)}{2(k-q+1)(p+q-k)}} \\
 d_1   & = & \sqrt{\frac{(k_1+2)(k_1-q_1+1)(p_1+q_1-k_1)
(2I_1+1)}{(I_1+1)(I_1+I_2+I+2)(I_1-I_2+I+1)}} \\
d_2    & = & \sqrt{\frac{(k_2+2)(k_2-q_2+1)(p_2+q_2-k_2)
(-I_1+I_2+I+1)}
{(2I_2+1)(I_2+1)(I_1+I_2+I+2)}} \\
d_3    & = & \sqrt{\frac{(l_2+1)(q_2-l_2)(p_2+q_2-l_2+1)(I_1+I_2-I)}
{I_2(2I_2+1)(I_1-I_2-I+1)}}.
\end{eqnarray*}
The second of these two expressions is not the most such general relation
which can be derived, but when used in combination with the first, it is
sufficient to determine the value of any ISF for the given irrep coupling,
once the values of the ISF's for $k,l = SHW$ are known.

\medskip\medskip
{\flushleft {\bf IV. Determining the isoscalar factors}}
\medskip

To move from the recursion relations to determination of the ISF, a sign
convention and a resolution scheme for outer degeneracy must be chosen.
For a given coupling,
\[ (p_1,q_1) \otimes (p_2,q_2) \rightarrow (p,q) \]
the degeneracy is determined by equation~(\ref{degene}).  For cases of
degeneracy $\eta = 1$, the choice of sign of one of the nonvanishing
ISF's for $(k,l) = SHW$ is sufficient to determine all the others.
In practice, one such ISF is set equal to 1; equations (\ref{RR1})
and (\ref{RR2}) are used to generate all others
from this one; and all are multiplied by a common factor to enforce
the normalization condition
\begin{equation}
   \sum_{k_1,l_1,k_2,l_2} \left( F(SHW:k_1,l_1;k_2,l_2) \right)^2 = 1.
\end{equation}

Even in such simple cases, the particular ISF to initialize must be chosen
as one which allows use of the recursion relations (\ref{RR1}) and
(\ref{RR2}) to determine neighboring values, and a recursive path from
the starting point to arbitrary ISF's must be deduced.

When the $(k,l)=SHW$ ISF's are all known, equation (\ref{RR4}) can be
used to deduce all \linebreak $k<p+q, l=0$ ISF's, and from them equation
(\ref{RR3}) implies all $l>0$ cases.

The most delicate problem is the determination of an algorithm which
uniquely determines all ISF's in cases of outer degeneracy two or higher.
Such an algorithm has been developed and implemented in C language codes
for evaluation of arbitrary ISF's as floating point values, and as exact
precision square roots of a ratio of
integers \cite{K&W}.  The ISF's produced by this algorithm share all the
symmetries under irrep exchange and conjugation with the familiar SU2
Clebsch-Gordan coefficients.

The logic of the algorithm can be made clearer through a change of variables.
Of the four integers, $k_1,l_1,k_2,l_2$, used heretofore as parameters of the
isoscalar factor for a coupling to a state of highest weight, only three are
independent.  The hypercharge conservation relation (\ref{hyperc}) implies
\[ k_1+l_1+k_2+l_2 = \frac{1}{3} (2(p_1+p_2)+4(q_1+q_2) +p-q). \]
Use of the definition
\[ s \equiv k_1-l_1+k_2-l_2 \]
allows the ISF's to be expressed as $F^n(SHW: s,k1,l1)$.   The degeneracy
index $n = 0, 1, \ldots \eta-1$ is necessary for couplings with degeneracy
$\eta >1$.  The algorithm works as follows:
\begin{itemize}
   \item make the following assignments for $0 \leq n < \eta$, $0 \leq
 n^{\prime} < \eta$

  $F^{n}(SHW: s_{max}-2n^{\prime},k_{1 \, min},l_{1 \, min}) =
 {\delta}_{n,n^{\prime}}$
where $\delta$ is the Kroneker delta.
   \item using these assignments, the recursion relations (\ref{RR1}) and
(\ref{RR2}) are adequate to determine all ISF's (with $k,l = SHW$) for
$s_{max} \geq s \geq s_{max} - 2(\eta - 1).$
   \item ISF's (with $k,l = SHW$) for remaining values of $s$ can be
determined without further assignments, evaluating each set of values
for fixed $s$ before moving to lower $s$.  To move to a lower $s$ value,
the recursion relation reduces to only three terms either at
$k_1 = k_{1 \, max}$, $l_1 = l_{1 \, min}$; or at $k_1 = k_{1 \, min}$,
$l_1 = l_{1 \, max}$.  This allows determination of one ISF for the new
$s$ value, which then allows the evaluation of all others at this $s$
value using the full four-term recursion relation.
This stepdown procedure fails in a small
subset of cases, whereupon one must take advantage of permutation symmetry
(discussed below) to move to the lower value of $s$.
   \item  The $F^0(SHW: s,k_1,k_2)$ values, once normalized, are proper
isoscalar factors.  A linear combination of the $F^0$'s and the $F^1$'s,
made orthogonal to $F^0$ and normalized, become proper $F^1(SHW: s,k_1,k_2)$
values.  Likewise, using the Gram-Schmidt orthogonalization procedure,
each set of $F^n$ with higher $n$ is
constructed from those with lower $n$'s.
   \item Remaining ISF's for values of $k,l \neq SHW$ can be determined in
a straightforward way using the remaining recursion relations, (\ref{RR3})
and (\ref{RR4}).
\end{itemize}

In this description, $s_{max}$ is the maximum value of $s$ for which a
non-vanishing ISF occurs for a coupling of fixed $p,q,p_1,q_1,p_2,q_2$
values:  $k_{1 \, min}$, $l_{1 \, min}$,$k_{1 \, max}$,$l_{1 \, max}$
are the minimum and maximum values of the $k_1$ and $l_1$ variables for
a particular value of $s$.

\medskip\medskip
{\flushleft {\bf V. Symmetries of the isoscalar factor}}
\medskip

The symmetries of the SU2 Clebsch-Gordan coefficient under permutation
of irreps ($j_i,m_i$) and under conjugation ($j,m \rightarrow j,-m$),
known as the Racah symmetries, are well known and frequently utilized
to simplify tabulations of coefficients and
recoupling calculations.  For SU3 couplings of degeneracy one, a complete
set of Racah symmetries can be demonstrated.  The algorithm described in
the previous section extends these symmetries to couplings of arbitrary
degeneracy, in contrast to some other degeneracy resolution schemes.
Once the symmetries of the SU3 WCG are
known, one can use the symmetry relations for the SU2 Clebsch-Gordan
coefficients to deduce symmetry properties of the SU3 isoscalar factors.

Derivation of the symmetry relations for the SU3 WCG is straightforward,
albeit tedious.  If one applies the $V_+$ operator (equation (\ref{vplus}))
to both sides of the defining expression for the WCG (equation
(\ref{wcgdef})), the result is a linear expression involving
six WCG's of various indices and corresponding
coefficients which sum to zero.  Under each of the transformations
\begin{enumerate}
   \item $(p_1,q_1; k_1,l_1,m_1) \leftrightarrow (p_2,q_2; k_2,l_2,m_2)$,
referred to as $(1 \leftrightarrow 2)$;
   \item $(p_1,q_1; k_1,l_1,m_1) \rightarrow (q,p; p+q-l,p+q-k, p+q-m)$ and
$(p,q; k,l,m) \rightarrow (q1,p1; p1+q1-l1, p1+q1-k1, p1+q1-m1)$, referred
to as $(1 \leftrightarrow \tilde{3})$; and
   \item $(p,q; k,l,m) \rightarrow (q,p; p+q-l,p+q-k, p+q-m)$ and similarly
for the states $p_1,q_1$ and $p_2,q_2$, referred to as {\it conjugation},
\end{enumerate}
the coefficients in the six term expression of WCG's transform among one
another in pairs, easily exhibiting the fact that the transformed WCG's
(within a sign which depends on the $k,l,m$ indices) obey the same six
term recursion relations as the original WCG's.

The symmetry transformations can also involve a sign change which depends
upon the $p$ and $q$ variables.  This is fully dependent upon another sign
convention which must be chosen in order to fully specify the WCG's and
ISF's.  Convenient choices are:
\[ \cgc{j_1}{j_2}{j}{m_1 = j_1}{m_2 = - j_2}{} > 0 \]
the familiar Condon and Shortley phase convention \cite{C&S}, and
\[ F^n(SHW: SHW_1; k_{2 \, max}, l_{2 \, min}) > 0. \]
with the $s$ value of the $F$ chosen to be positive given as $s_{max} - n$.
To derive the $p,q$ dependence of the phase under one of the symmetry
transformations, begin with a SU3 WCG which is positive under this
convention; apply the transformation; and determine the sign of the
transformed WCG relative to that which is positive
 by convention among the transformed coefficients, using the recursion
relations derived earlier.  This sign, which will depend upon the six
$p,q$ values only, becomes part of the symmetry relation.

The absolute magnitude of the ratio of a WCG to its permuted version
results from the normalization condition
\begin{equation}
    \sum_{\kappa_1, \kappa_2} (\wcg{{\cal P}_1}{{\cal P}_2}{{\cal P}}{n}
{\kappa_1}{\kappa_2}{\kappa})^2 = 1.
\end{equation}
It is straightforward to show that two of the transformations --
 $(1 \leftrightarrow 2)$ and conjugation -- produce no change in
normalization and thus require no constant term; and the third --
 $(1 \leftrightarrow \tilde{3})$  -- requires a constant equal to the
square root of the ratio of the dimension of the irreps
${\cal P}_1$ and ${\cal P}$.

When one considers couplings which have degeneracy greater than one,
each set of distinct WCG's (labeled by the index {\it n}) has a unique
element chosen as positive by convention.  As a result, there is an
additional phase contribution of $(-1)^n$ in each of the transformations
considered here.

The resulting symmetry relations are as follows:
\begin{equation}
   \wcg{{\cal P}_1}{{\cal P}_2}{{\cal P}}{n} {\kappa_1}{\kappa_2}{\kappa}
 = (-1)^{\gamma + \sigma + max(\gamma, \sigma)+n} \; \wcg{{\cal P}_2}
{{\cal P}_1}{{\cal P}}{n} {\kappa_1}{\kappa_2}{\kappa}
\end{equation}
for the $(1 \leftrightarrow 2)$ transformation;
\begin{equation}
   \wcg{{\cal P}_1}{{\cal P}_2}{{\cal P}}{n} {\kappa_1}{\kappa_2}{\kappa}
 = (-1)^{m_2+n} \, \sqrt{ \frac{(p+1)(q+1)(p+q+2)}{(p_1+1)(q_1+1)
(p_1+q_1+2)}} \, \wcg{\tilde{\cal P}}{{\cal P}_2}{{\tilde{\cal P}}_1}{n}
 {\tilde{\kappa}}{\kappa_2}{\tilde{\kappa}_1}
\end{equation}
for the $(1 \leftrightarrow \tilde{3})$ transformation (where
$\tilde{\cal P}$ represents the ordered pair $q,p$ and $\tilde{\kappa}$
represents $p+q-l, p+q-k, p+q-m$); and
\begin{equation}
   \wcg{{\cal P}_1}{{\cal P}_2}{{\cal P}}{n} {\kappa_1}{\kappa_2}{\kappa}
 = (-1)^{\gamma + \sigma + min(\gamma,\sigma)+n} \, \wcg{\tilde{{\cal P}}_1}
{\tilde{{\cal P}}_2}{\tilde{{\cal P}}}{n} {\tilde{\kappa}_1}{\tilde{\kappa}_2}
{\tilde{\kappa}}
\end{equation}
for the conjugation transformation.

No additional information is given by repeating this procedure with the
ladder operators ${\hat{V}}_-$ and ${\hat{U}}_+$ since they are related
to the two operators already considered by Hermitian conjugation:  similar
treatment using ${\hat{T}}_+$ and ${\hat{T}}_-$ generate the symmetry
relations for the SU2 Clebsch-Gordan coefficients.

Corresponding symmetry relations for the isoscalar factors follow from
the above combined with the symmetry relations for the SU2 Clebsch-Gordan
coefficients in the Condon and Shortley phase conventions.  They are
\begin{equation}
  F( {\cal P}, \kappa: {\cal P}_1, {\kappa}_1; {\cal P}_2, {\kappa}_2)
   = (-1)^{\gamma + \sigma + max(\gamma,\sigma) + I_1 +I_2 - I} \,
 F( {\cal P}, \kappa: {\cal P}_2, {\kappa}_2; {\cal P}_1, {\kappa}_1)
\end{equation}
for the $(1 \leftrightarrow 2)$ transformation (here $\kappa$ represents
the pair $(k,l)$;
\begin{eqnarray}
  F( {\cal P}, \kappa: {\cal P}_1, {\kappa}_1; {\cal P}_2, {\kappa}_2)
  &=&(-1)^{l_2} \,
   \sqrt{ \frac{(p+1)(q+1)(p+q+2)(k_1-l_1+1)}{(p_1+1)(q_1+1)(p_1+q_1+2)
(k-l+1)}}
\nonumber \\ && F( \tilde{{\cal P}}_1, \tilde{\kappa}_1: \tilde{{\cal P}},
\tilde{{\kappa}};  {\cal P}_2, {\kappa}_2)
\end{eqnarray}
for the $(1 \leftrightarrow \tilde{3})$ transformation; and
\begin{equation}
  F( {\cal P}, \kappa: {\cal P}_1, {\kappa}_1; {\cal P}_2, {\kappa}_2)
   = (-1)^{\gamma + \sigma + max(\gamma,\sigma) + I_1 +I_2 - I} \,
  F( \tilde{\cal P}, \tilde{\kappa}: \tilde{\cal P}_1, \tilde{\kappa}_1;
 \tilde{\cal P}_2, \tilde{\kappa}_2)
\end{equation}
for the conjugation transformation.

The choice of a resolution procedure which yields symmetries such as these
produces considerable simplifications in calculations of physical states,
reduces the complexity of definitions of $6-j$ and $9-j$ type recoupling
coefficients, and significantly shortens databases and
printed tables of WCG's and ISF's.

\medskip\medskip
{\flushleft {\bf Acknowledgements}}
\medskip

The author would like to thank L. C. Biedenharn for helpful discussions
of the problems work described herein; M. Danos for reading an early
version of this manuscript and many useful suggestions; and T. A. Kaeding
whose work in implementing the algorithm
 in explicit code has corrected several errors and helped refine the
algorithm.

\end{document}